\documentclass[preprintnumbers, prb, showpacs, floatfix,twocolumn,
preprintnumbers, letterpaper, superscriptaddress]{revtex4}
\usepackage{amsfonts}
\usepackage{amsmath}
\usepackage{amssymb,epsf}
\usepackage{latexsym}
\usepackage{graphicx,epsfig}
\usepackage{amssymb}
\usepackage{subfigure}
\usepackage[colorlinks=true,citecolor=blue,linkcolor=blue,urlcolor=black]{hyperref}
\usepackage{epstopdf}
\usepackage{color}

\def\be{\begin{equation}}
\def\ee{\end{equation}}
\def\ba{\begin{eqnarray}}
\def\ea{\end{eqnarray}}
\def\la{\langle}
\def\ra{\rangle}

\begin{document}
\title{ Classical criticality establishes quantum topological order}
\author{Mohammad Hossein Zarei}
\email{mzarei92@shirazu.ac.ir}
\affiliation{Physics Department, College of Sciences, Shiraz University, Shiraz 71454, Iran}
\author{Afshin Montakhab}
\email{montakhab@shirazu.ac.ir}
\affiliation{Physics Department, College of Sciences, Shiraz University, Shiraz 71454,
Iran}

\begin{abstract}
We establish an important duality correspondence between
topological order in quantum many-body systems and criticality in
ferromagnetic classical spin systems. We show how such a
correspondence leads to a classical and simple procedure for
characterization of topological order in an important set of
quantum entangled states, namely the Calderbank-Shor-Steane (CSS)
states. To this end, we introduce a particular quantum Hamiltonian
which allows us to consider the existence of a topological phase
transition from quantum CSS states to a magnetized state. We study
the ground state fidelity in order to find non-analyticity in the
wave function as a signature of a topological phase transition.
Since hypergraphs can be used to map \emph{any} arbitrary CSS
state to a classical spin model, we show that fidelity of the
quantum model defined on a hypergraph $H$ is mapped to the heat
capacity of the classical spin model defined on dual hypergraph
$\tilde{H}$. Consequently, we show that a
ferromagnetic-paramagnetic phase transition in a classical model
is mapped to a topological phase transition in the corresponding
quantum model. We also show that magnetization does not behave as
a local order parameter at the transition point while the classical order parameter is
mapped to a non-local measure on the quantum side, further indicating
the non-local nature of the transition. Our procedure not only
opens the door for identification of topological phases via the
existence of a local and classical quantity, i.e. critical point,
but also offers the potential to classify various topological
phases through the concept of universality in phase transitions.
\end{abstract}
 \maketitle
\section{Introduction}
One of the most important problems in condensed matter physics is
to characterize the different phases of matter which are related
to various kinds of order present in physical systems. The quantum
counterpart of this problem is in particular interesting and has
attracted much attention over the past decades \cite{sav}. While most
well-known orders are classified based on symmetry properties of
the system, there are physical systems with a topological order
\cite{20, 23, 21, 22,wen,t1,t2,t3}, associated with topological
properties of the system instead of symmetries. In particular,
topological order has a non-local nature in a sense that there is
no local order parameter to characterize a topological phase
transition under a symmetry-breaking mechanism. Consequently,
unlike symmetry-breaking phases, characterizing topological phases
has remained a challenging problem to this date \cite{wwen,wang,
wen3,wen4}.

Furthermore, topological order, due to its non-local nature, has been also an important concept in quantum information theory.
It is specifically important in quantum error correction as a way to overcome the
decoherence problem in quantum computers. Using general quantum error-correcting codes,
specifically Calderbank-Shor-Steane (CSS) codes \cite{Calderbank1996,stean,gottesman,cssd}, one usually needs an active protocol for error
correction. However, due to topological nature of certain CSS
codes, they exhibit self-correction \cite{alicki,bom,pach,errork},
thus protecting information in a natural way. Generally,
topological nature of various CSS codes makes them an immediate
candidate for fault-tolerant quantum computing
\cite{Kitaev2003,bombin2006,yoshida, bombin2015,rob2,
zare16}. Therefore, being able to ascertain whether an arbitrary
CSS code is topological or not is of fundamental importance
particularly in their feasibility as quantum memory. One
standard method is to identify measures such as topological
entanglement entropy \cite{kitpre}, or gap stability
\cite{bravyi}. While such measures can well capture certain
physical aspects of topological order, they nonetheless have a
non-local nature. Therefore, a practical procedure for
characterizing topological order for an arbitrary CSS quantum state by
such measures remains a challenging problem \cite{nat}.

On the other hand, during the past decade certain interesting maps
from quantum entangled states, specially topological ones, to
partition functions of classical spin models have been introduced
\cite{castel2005, Nest2007, zarei18} which have led to new
developments in quantum information theory as well as statistical
mechanics \cite{Geraci2008,Geraci2010,Somma2007,Verstraete2006,Nest2008,Vahidb}.
In the light of such mappings, one may explore the consequences of
the existence of a phase transition in the classical spin models
for the corresponding quantum entangled states.  In particular the
non-analytic behavior associated with classical phase transitions
should have important ramifications on the quantum side
\cite{zarei19, zareiab}.

In this paper, we propose and prove that the existence of
topological order in an arbitrary CSS state is identified by a
critical ferromagnetic phase transition in their classical (dual)
spin model.  We therefore propose a simple mathematical procedure:
criticality in the classical partition function establishes
topological order in the corresponding CSS states. This is
important since it provides a classical local measure for a
quantum non-local phenomenon. In order to prove this, we propose a
quantum Hamiltonian model, the CSS-magnetic (CSSM) model, which
exhibits a transition as a function of control parameter
($\beta$), from the magnetized state to a CSS state. Due to the
specific property of the ground state, we show that it cannot
break any (spin) symmetries. We next show that the transition to
the CSS state is indeed characterized by a singularity in the
ground state fidelity and thus a quantum phase transition. We
therefore conclude that the quantum phase transition must be of
topological nature since it preserves the symmetry. As further
evidence for the topological nature of the quantum phase
transition, we also calculate the magnetization and show that it
does not behave as a local (symmetry-breaking) order parameter,
while the classical order parameter behaves as a non-local
measure on the quantum side. In
effect, we provide a duality between a (nonlocal) quantum
topological phase transition and a (local) classical ferromagnetic
phase transition.

\section{CSS-magnetic model on a hypergraph }
\begin{figure}[t]
\centering
\includegraphics[width=7cm,height=3cm,angle=0]{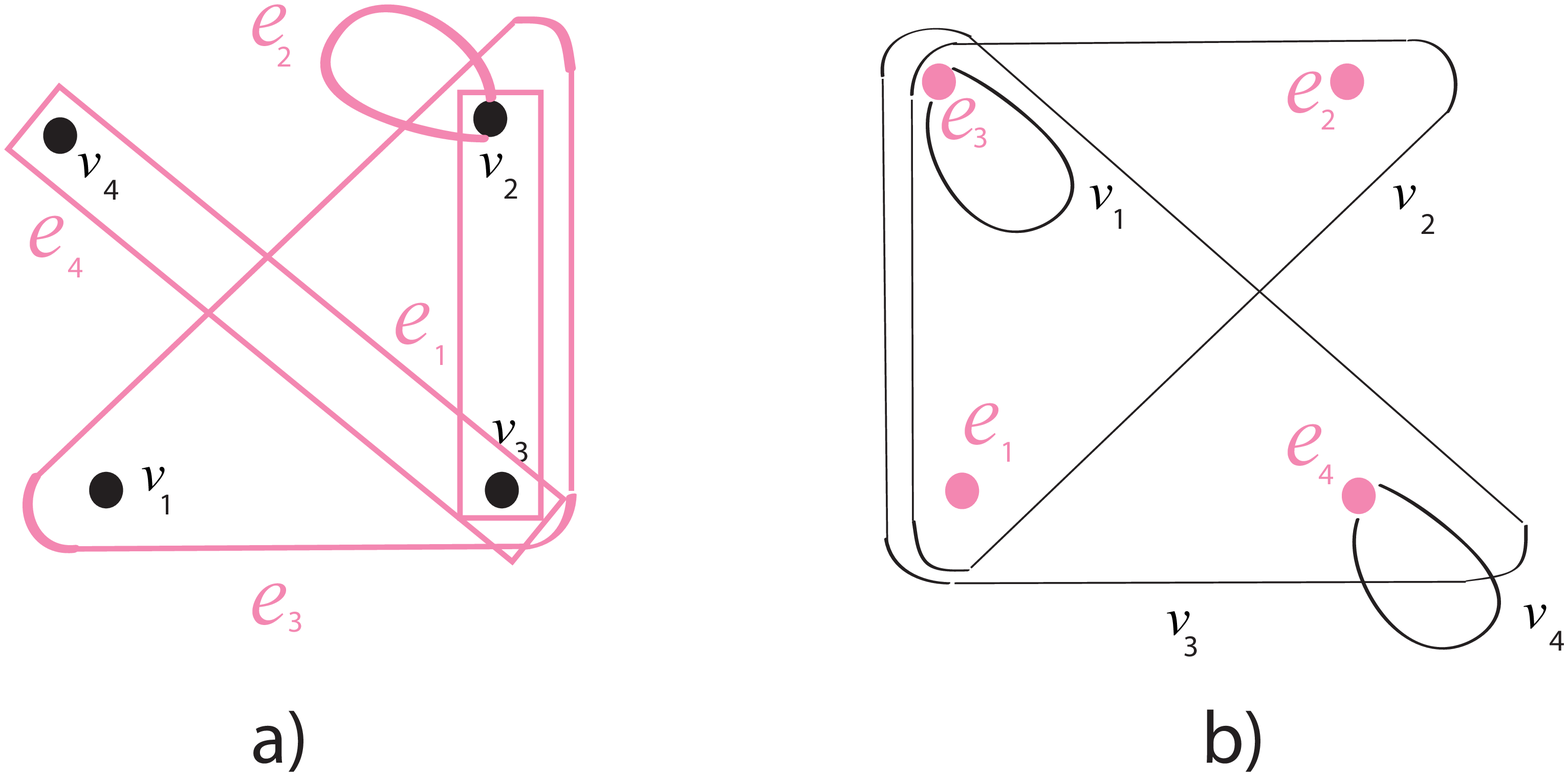}
\caption{a) A simple hypergraph including four vertices denoted by
blue (dark)circles and four hyperedges denoted by pink (light)
curves. b) Dual hypergraph of part (a).} \label{hyp}
\end{figure}

 A stabilizer state
on $N$ qubits is a positive eigenstate of $N$ commutative
operators belonging to the Pauli group. A CSS state is a
particular set of stabilizer states which is stabilized by
$Z$-type and $X$-type operators \cite{gottesman}. CSS states can be defined on hypergraphs
\cite{zareiprb,qhyp1, qhyp2}. A hypergraph $H=(V,E) $ is a set of
vertices $V=\{v_1 , v_2 , ..., v_K \}$  and a set of hyperedges
$E=\{e_1 , e_2 , ..., e_N \}$ where each hyperedge is equal to a
subset of vertices, see Fig.(\ref{hyp}-a). A set of hyperedges are called independent if no
hyperedge is equal to collection of other hyperedges. For each
hepergraph $H=(V,E)$, one can define a dual hypergraph
$\tilde{H}=(\tilde{V},\tilde{E})$ where $\tilde{V}=\{\tilde{v}_1
,\tilde{v}_2 ,...,\tilde{v}_N \}$ and $\tilde{E}=\{\tilde{e}_1 ,
\tilde{e}_2 , ..., \tilde{e}_K\}$ are vertices and hyperedges of
$\tilde{H}$, respectively, and
\begin{equation}
\tilde{e}_i=\{\tilde{v}_m
| v_i \in e_m~in~ H\}
\end{equation}
where $v_i \in e_m~in~ H$ refers to vertices belonging to the edge
of $e_m$ on $H$.  Simply put, duality interchanges vertices and
edges, see Fig.(\ref{hyp}-b).

For a given hypergraph $H=(V,E)$, there is a CSS state in the
following form:
\begin{equation}\label{css}
|CSS_H\ra=\frac{1}{2^{\frac{M}{2}}}\prod_{e\in E}(1+A_e )|0\ra ^{\otimes N}
\end{equation}
where $|0\ra$ is the positive eigenstate of the Pauli operator $Z$
with $N$ qubits living on vertices and $M \leq N$ is the number of
independent hyperedges. $A_e$ is an $X$-type stabilizer in the
form of $\prod_{i\in e}X_i$ where $i\in e$ refers to all vertices
belonging to $e$. Furthermore, there are $K=N-M$ number of
$Z$-type stabilizers for the above state which commute with
$A_e$ \cite{zarei18}. A set of hyperedges corresponding to such $Z$-type
stabilizers are defined as orthogonal hyperedges $E^{*}$ where the CSS state of Eq.(\ref{css}) can
also be written in the following form:
\begin{equation}\label{css2}
|CSS_H\ra=\frac{1}{2^{\frac{K}{2}}}\prod_{e^{*}\in E^{*}}(1+B_{e^{*}} )|+\ra ^{\otimes N}
\end{equation}
where $|+\ra$ is the positive eigenstate of the Pauli operator
$X$. $e^{*}$ is a member of set of orthogonal hyperedges $E^{*}$
and $B_{e{*}}$ is a $Z$-type stabilizer in the form of
$\prod_{i\in e^{*}}Z_i$ where $i\in e^{*}$ refers to vertices
belonging to $e^{*}$. Furthermore, it is simple to check that the
above CSS state is also a non-degenerate ground state of a quantum
CSS model with a Hamiltonian in the form of $h=-\sum_{e^{*} \in
E^{*}}B_{e^{*}} -\sum_{e\in E}A_e$. We should note that while we
can choose the $X$-type stabilizers as local operators, some
$Z$-type stabilizers might be non-local. Therefore, the above
Hamiltonian might be regarded as non-physical. However, here we
are only concerned with existence of topological order in the
ground state of such Hamiltonian, namely the CSS state,  and not
whether such Hamiltonian actually represents any physical system. In other words, corresponding to each CSS state we can construct a physical Hamiltonian by removing non-local operators from the Hamiltonian of $h$ in a sense that there will be a degenerate ground space and the above CSS state will be only one of the ground states of the physical Hamiltonian. However, since topological order is a property of the wave function, we expect to see topological properties of the model in the above CSS state. Therefore, existence of topological order in such state means that its corresponding physical Hamiltonian will be a topological model with a topological degeneracy in the ground state.

We introduce an extended version of a CSS model, the
CSS-magnetic (CSSM) model, corresponding to a given hypergraph
$H=(V,E)$ in the following form:
\begin{equation}\label{qm}
\mathcal{H}=-\sum_{e^{*} \in E^{*}}B_{e^{*}} -\sum_{e\in E}A_e +\sum_{e\in E} U_e (\beta)
\end{equation}
where $U_e (\beta) = \prod_{i \in e} \exp\{-\beta Z_i \}$ is a
product operator corresponding to a hyperedge  $e$ with a tuning
parameter $\beta$. If we expand this operator for a small value of
$\beta$, it will correspond to a magnetic term in the first-order
approximation. Additionally,  the first two terms in the
Eq.(\ref{qm}) are stabilizers of a CSS state. Therefore, it is
easy to see that the ground state of the above Hamiltonian goes
through a transition from a CSS state ($\beta =0$) to a magnetized
state, $|0\ra^{\otimes N}$,  for $\beta\rightarrow \infty$.  In
this way, an important problem that needs be considered is the
possibility of a quantum phase transition in this model. It is
therefore important to identify the symmetry properties of the
ground state of the CSSM model and ask if such ground state
exhibits any singular behavior as a function of $\beta$.

To find the ground state, we re-write the Hamiltonian in
Eq.(\ref{qm}) in the following form:
\begin{equation}\label{qm1}
\mathcal{H}=-\sum_{e* \in E*}B_{e*} + \sum_{e\in E} Q_e (\beta)
\end{equation}
where $Q_e (\beta) = U_e -A_e$ which is a positive operator. One
can check that $Q_e ^2 =2\cosh (\beta \sum_{i}Z_i) Q_e$ and since
$\la Q_e ^2\ra \geq 0$, one concludes that $\la Q_e \ra \geq 0$
for any arbitrary quantum state. This implies that the minimum
eigenvalue of $Q_e $ is zero.

On the other hand, since $A_e $ and $U_e$ commute with $B_{e{*}}$,
it is clear that $[Q_e , B_{e^{*}}]=0$, and therefore, an
eigenstate of $Q_e $ corresponding to eigenvalue of zero will be
the ground state of the CSSM. One can easily find the zero-eigen state in the following form:
\begin{equation}\label{GH}
|G_H (\beta) \ra
=\frac{1}{\sqrt{\mathcal{Z}(\beta)}}\exp\{\frac{\beta}{2}\sum_{i}Z_i\}|CSS_H\ra
\end{equation}
where $\mathcal{Z}(\beta)$ is the normalization factor and
$\sum_{i} $ in the exponential term refers to summation on all
qubits. In order to show that the above state is in fact the
ground state, first note that $U_e |G_H (\beta)\ra
=\frac{1}{\mathcal{Z}(\beta)}\exp\{-\frac{\beta}{2}\sum_{i\in
e}Z_i\}\exp\{\frac{\beta}{2}\sum_{i ~\not\in  e}Z_i\}|CSS_H\ra$.
Then, since $A_e \exp\{\frac{\beta}{2}\sum_{i\in e}Z_i \}
=\exp\{-\frac{\beta}{2}\sum_{i\in e}Z_i \} A_e$, it follows that $
A_e |G_H (\beta)\ra =U_e |G_H (\beta)\ra$ and therefore $Q_e |G_H
(\beta)\ra =0$.

It remains to find the exact form of the normalization factor. We
now show that $\mathcal{Z}(\beta)$ is the partition function of a
classical ferromagnetic spin model defined on dual hypergraph
$\tilde{H}$ with the following classical Hamiltonian:
\begin{equation}
\mathcal{H}_{cl} =-J\sum_{\tilde{e}\in \tilde{E}}\prod_{i\in \tilde{e}}s_i
\end{equation}
where $J$ is the ferromagnetic coupling constant and $\prod_{i\in
\tilde{e}}s_i$ refers to many-body interaction between binary
spins belonging to a hyperedge $\tilde{e}$. Now, using
Eq.~(\ref{GH}), we have:
\begin{equation}\label{part}
\mathcal{Z}(\beta)=\la CSS_H |\exp\{\beta \sum_{i}Z_i\}|CSS_H\ra
\end{equation}
On the other hand, according to Eq.(\ref{css2}), the CSS state can
be written in terms of $Z$-type operators, and since $B_{e*}$
operators commute with $\exp\{\beta \sum_{i}Z_i\}$ and all
$B_{e*}$'s stabilize the CSS state, we will arrive at the
following form for Eq.(\ref{part}):
\begin{equation}\label{parti}
\mathcal{Z}(\beta)= 2^{\frac{K}{2}}\la \alpha |CSS_H\ra
\end{equation}
where $|\alpha \ra =\exp\{\beta \sum_{i}Z_i\} |+\ra ^{\otimes
N}=2^{-\frac{N}{2}}(e^{\beta}|0\ra +e^{-\beta}|1\ra)^{\otimes N}$.
Indeed, such a product-state of a CSS state defined on a
hypergraph $H$ is equal to the partition function of a classical
ferromagnetic spin model defined on dual hypergraph $\tilde{H}$ up
to a factor $2^{\frac{K}{2}-N}$, with $\beta$  being related to
the temperature in the classical model in the form of
$T=\frac{J}{k_B \beta}$, as has been shown in \cite{zarei18}. In
the following we set $k_B$ and $J$ to unity so that
$T=\frac{1}{\beta}$.

Finally, we note that relation of the normalization factor in
our model to the partition function of a classical model is in
fact a result of specific direction of the quantum state. In other
words, the unnormalized quantum state is in the form of $\exp\{\frac{\beta}{2}\sum_{i}Z_i\}|CSS_H\ra$. It is simple to check that if one expands
the state $|CSS_H \ra$ in the computational basis of $|0\ra$ and
$|1\ra$ and then applies the operator  $\exp\{\frac{\beta}{2}\sum_{i}Z_i\}$,
the above state will be a superposition of computational bases
with weights which are related to the Boltzmann weights
corresponding to different configurations of a classical spin
model. It is exactly this reason that the normalization factor is
mapped to the partition function as it is a summation of the
Boltzmann weights.

Obtaining the exact form of the ground state, we are ready to
consider the existence of a quantum phase transition in the CSSM.
First, we consider the symmetries of the ground state in order to
address the possibility of a symmetry breaking phase transition.
We first consider the $\beta \rightarrow \infty$ limit where one
can see that the ground state is a magnetized state, $|00...0\ra$.
It is clear that, in this extreme, any $Z$-type operator is a
symmetry operator of the ground state. However, when we decrease
$\beta$ to a finite value, only the $Z$-type operators which are
stabilizers of the CSS state, i.e. $B_{e^{*}}$, remain as the
symmetries of the ground state, independent of the \emph{finite}
value of $\beta$. Therefore, the only symmetry-breaking transition
that might occur in the ground state must occur at
$\beta\rightarrow\infty$, i.e. at zero temperature. All other
finite $\beta$ ground states of the CSSM model possess the
symmetries of $B_{e^{*}}$ operator. We conclude that any phase
transition occurring at finite $\beta$ cannot be a symmetry
breaking transition and thus must be a topological phase
transition. In fact, quantum topological phase transition is
typically accompanied by long-range entanglement without any
symmetry breaking property. The signature of such transitions are
encoded in the ground state of the system and are usually studied
using tools of quantum information theory such as measures of
entanglement as well as fidelity \cite{hamma, gu, chen, fidelity,fisher,uhlmann,castel,mont2010}.
Ground state fidelity as a function of $\beta$ can encode such a
transition and thus show a singular behavior.
We next calculate such a quantity and show that is is equivalent to
the heat capacity of the classical spin model which exhibits a singular behavior
at the ferromagnetic phase transition at finite $\beta$, thus establishing the corresponding
topological phase transition in the CSSM model.

\section{Ground state fidelity}
 The ground state fidelity will
be a function of $\beta$ and $\delta \beta$. However, since
$\delta \beta$ is a very small quantity we can expand fidelity in
terms of $\delta \beta$. We first consider the exact form of
ground state fidelity as follows:
\begin{equation}
F=\frac{ \la CSS_H |\exp\{(\beta+\frac{\delta \beta }{2})\sum_{i}
Z_i\}|CSS_H\ra}{\sqrt{\mathcal{Z}(\beta)\mathcal{Z}(\beta + \delta
\beta)}}.
\end{equation}
Since the inner product term in the above relation is again
related to a partition function we will have:
\begin{equation}
F=\frac{\mathcal{Z}(\beta +\frac{\delta \beta}{2})}{\sqrt{\mathcal{Z}(\beta)\mathcal{Z}(\beta + \delta \beta)}}.
\end{equation}
After a Taylor expansion for the above equation, we find the
following form up to the second-order:
\begin{equation}
F(\beta ,\delta \beta) \simeq 1- \frac{1}{8} (\frac{\partial ^2
\ln(\mathcal{Z})}{\partial \beta^2})\delta \beta ^2
\end{equation}
where $\mathcal{Z}(\beta)$ is the partition function of a
classical spin model on the hypergraph $\tilde{H}$. On the other
hand, heat capacity of a classical spin model is given by  $C_v
=\frac{1}{T^2}\frac{\partial^2 \ln(\mathcal{Z})}{\partial^2
\beta}$ where $\beta$ is the inverse temperature, $\beta=1/T$.
Therefore, we have shown that the ground state fidelity of the
CSSM on a hypergraph $H$ is related to the heat capacity of a
classical spin model on dual hypergraph $\tilde{H}$:
\begin{equation}\label{fid}
F_H (\beta ,\delta \beta)\simeq1-\frac{(C_v)_{\tilde{H}}
}{8 \beta^2}\delta \beta ^2
\end{equation}
Now, it is known that if a classical spin model has a second-order
phase transition at a critical temperature $T_c$, the heat
capacity shows a singularity at $T_c$ where it diverges according
to $C_v\sim(T-T_c)^{-\alpha}$ where $\alpha$ is a critical
exponent of the classical spin model \cite{path}. We conclude that
the ground state fidelity of a CSSM on hypergraph $H$ shows a
singularity at a critical value of $\beta_c$ if its classical dual
shows a phase transition at a finite temperature of
$T_c=1/\beta_c$. Therefore, a second-order phase transition in a
ferromagnetic classical spin model establishes a quantum phase
transition in the corresponding CSSM model, and since any finite
$\beta$ transition is a non-symmetry breaking transition, it must
therefore be a topological phase transition, and thus indicating
the existence of a topological phase in the CSS state. We should
emphasize that such a procedure for identification of a
topological phase transition works for any arbitrary CSS state. In
the Appendix (\ref{apen}), we have provided two particular
examples in order to clarify the usefulness of the procedure
further.

\section{Order parameter considerations}
 Since CSS states with a well-defined thermodynamics limit are
classified as gapped quantum liquids \cite{wen4}, and since such
systems are known to have quantum phase transitions which are
categorized as symmetry-breaking, first-order, trivial, or
topological, we see that our method leads us to conclude that our
quantum phase transition is indeed a topological one. However, we
need to prove that our consideration of $Z_2$ symmetry is
sufficient for the purpose of establishing a topological phase. To
this end, we look for an indicator of topological phase transition
in the CSSM in a form of an order parameter. First, let us
consider the magnetization of the CSSM, defined as $m_H =\frac{\la
\sum_{i}Z_i \ra}{N}$, which can be written as:
\begin{equation}
m_H = \frac{\la CSS_H |(\sum_{i} Z_i )\exp\{\beta
\sum_{i}Z_i\}|CSS_H \ra}{ N\mathcal{Z}(\beta)}
\end{equation}
We now replace the operator $\sum_{i} Z_i$ with a derivative of
$\exp\{\beta \sum_{i}Z_i\}$. Then, the magnetization of the CSSM
is related to the internal energy of a classical ferromagnetic
spin model in the following form:
\begin{equation}
m_H =\frac{1}{N} \frac{1}{\mathcal{Z}(\beta)}\frac{\partial \mathcal{Z}}{\partial \beta} =- \frac{1}{N} \mathcal{E} _{\tilde{H}}
\end{equation}
where $\mathcal{E} _{\tilde{H}}$ refers to the internal energy of
a classical ferromagnetic spin model defined on the hypergraph
$\tilde{H}$. On the other hand, it is known that the internal
energy of a ferromagnetic model does not behave like an order
parameter as it displays a gradual (smooth) transition from a
negative value to zero. Therefore, it is concluded that
magnetization of the CSSM cannot be an order parameter, a fact
that further implies the topological nature of the above
transition.

On the other hand, one might expect to find an order parameter for
the CSSM by considering the classical (local) order parameter,
$\la S_i \ra$. In the mapping from classical partition function to
CSS state, each multi-spin interaction of $\prod_{i\in e} S_i$
corresponding to the hyperedge $e$ is mapped to a Pauli operator
$Z_e$ \cite{zarei18}. Therefore, if we perform an inverse mapping,
each spin variable $S_i$ will be equal to a product of $Z_e$
operators. Therefore, the order parameter for the quantum phase
transition in the CSSM model will be the expectation value of
suitable products of $Z_e$ operators.  Note that while such order
parameter might be difficult to calculate, one can see that it
must exist due to the nature of the inverse mapping. Now if such a
quantum order parameter is local, it must possess $Z_2$ symmetry,
which we have argued is impossible for our model. Therefore it
must be non-local, thus proving that a non-local order parameter
exist, which is sufficient to establish the existence of a
topological phase.

\section{Discussions}
The fact that quantum entangled states can be mapped to partition
function of classical spin models has many important consequences.
One that has not attracted much attention in the literature is the
consequence of singularities associated with criticality on the
classical side.  In this paper we have taken a step in this
direction and have found that if the classical dual of a CSS state
displays a critical point at a finite temperature $T$, the CSS
state has topological order. We point out that evidences for such
a correspondence already exists \cite{zarei18}. For example, the
toric codes defined on arbitrary graphs are mapped to Ising models
with ferromagnetic phase transition. GHZ states which do not have
topological order are mapped to one dimensional Ising models which
do not show a phase transition. Another example are graph states
without any topological order which are mapped to Ising models in
presence of magnetic field which do not show a phase transition.
However, we have established that given an \emph{arbitrary} CSS
state, one can easily identify whether its classical counterpart
has a ferromagnetic phase transition and thus conclude that it
must have topological order. This procedure is simple and direct
because the existence of critical point is established either by
simple observation or, in more complicated case, by numerical
simulations. Two concrete examples of such a procedure are
discussed in the Appendix.

The problem of identifying topological phases is a challenging
open problem. Our work offers some new insights in this regard.
For example, new topological states may be found by (inverse)
mapping classically critical ferromagnetic models via hypergraphs.
Also, our procedure can be used to check some controversial
aspects of the topological nature of certain recently proposed CSS
states such as X-cube model \cite{xcub}. Furthermore, the fact
that symmetry-breaking phase transition can be used to identify a
symmetry preserving topological phase may have important
consequences. As pointed out above, the potential for finding a
non-local order parameter on the quantum side via mapping of
classical local order parameter offers an intriguing possibility.
We also note that our CSSM model belongs to the well-known family
of stochastic matrix form decomposition \cite{castel2005} where
their classical-quantum correspondence has been studied for some
two-dimensional models \cite{Verstraete2006}. Our results may
reveal other aspects of importance of such family of states for
studying topological properties of quantum states.

Another interesting possibility is the observation that fidelity
was mapped to the heat capacity whose singularity is characterized
by the exponent $\alpha$.  Standard statistical mechanics tells us
that the divergence of heat capacity (and other non-analytic
behavior) at the critical point are \emph{universal} in a sense
that they depend only on the symmetries of the classical
Hamiltonian, and not on the details of interactions, etc. Thus,
various different systems fall in the same universality class
displaying the same exponents. It is well-known that the scaling
of quantum measures such as fidelity at the quantum phase
transition point can be related to scaling of the correlation
length, and that the correlation length exponent can be used to
define universality classes \cite{rad18,chen17}.  Therefore, the
potential of applying the concept of universality, which is based
on local symmetries of the classical model, to classify various
topological phases offers an interesting prospect.

Finally, we note that the problem of topological phase transition
and its relation to a wider class of quantum phase transitions has
been studied by various authors before.  Here, we have found a
mechanism, by choosing a specific perturbation which preserves
symmetry, to map a non-local quantum phase transition to a local
classical phase transition, and consequently use this mechanism as
a diagnosis for the existence of topological phase in an important
class of quantum states.  Whether one can find similar mechanisms
to embrace a more general class of quantum states poses an
interesting possibility for future work.

\section*{Acknowledgement}
We would like to thank A. Ramezanpour for useful discussions
during preparation of this paper.

\section{Appendix}\label{apen}
In this Appendix, we consider two particular examples to show how
a well-known classical ferromagnetic-paramagnetic phase transition
can reveal a topological phase transition in a corresponding
quantum CSSM model. In particular, we consider classical Ising
models which have been well studied. In particular, since the
one-dimensional classical Ising model does not show a phase
transition at a finite temperature we expect that the
corresponding quantum model does not show a topological phase
transition. On the other hand, the two dimensional Ising model has
a finite temperature phase transition and there must be a
topological phase transition in the corresponding quantum model.

Let us start with the one-dimensional Ising model where classical
spins live in vertices of a one dimensional periodic lattice and
there is a two-body interaction between two neighboring spins.
According to hypergraph duality explained in Ref.\cite{zarei18},
this model is mapped to a GHZ state in the form of
$\frac{1}{\sqrt{2}}(|++...+\ra +|--...-\ra)$ which is stabilized
by $X$-type stabilizers in the form of $X_i X_{i+1}$ and a
non-local $Z$-type stabilizer in the form of $B =\prod_{i} Z_i$
where $\prod_i$ refers to all vertices of the lattice. Therefore,
the corresponding CSSM model, according to definition in
Eq.(\ref{qm}), will be described by the following Hamiltonian:
\begin{equation}
H = - B -\sum_i X_i X_{i+1} + \sum_i e^{-\beta (Z_i + Z_{i+1})}
\end{equation}
Note that it is similar to a quantum Ising model perturbed by a
term in the form of $e^{-\beta (Z_i + Z_{i+1})}$. In particular,
if one rewrites $e^{-\beta Z_i}$ in the form of $\cosh \beta
-\sinh \beta ~Z_i$, the Hamiltonian will find the following form:
$$H = \cosh^2 \beta - B - \sum_i X_i X_{i+1}$$
\begin{equation}\label{0}
 -(2 \sinh \beta ~ \cosh\beta) \sum_i  Z_i +(\sinh ^2 \beta) \sum_{i} Z_i Z_{i+1}
\end{equation}

It is clear that the above Hamiltonian is \emph{not} an Ising
model in a transverse field because of an additional term of
$\sum_i Z_i Z_j$. In particular, it is not \emph{a priori} clear
if the above Hamiltonian can show a quantum phase transition at a
finite value of $\beta$. However, our mapping can provide a
solution to the above problem because the quantum Hamiltonian is
mapped to a one-dimensional classical Ising model which has a
phase transition at zero temperature. Therefore, according to our
procedure,  we conclude that the above quantum Hamiltonian has a
phase transition point at infinite value of $\beta$.

On the other hand, our methodology can further explain the nature
of this phase transition. According to our logic the exact ground
state of the above quantum Hamiltonian will be in the following
form:
 \begin{equation}\label{1}
 |G (\beta)\ra =e^{ \frac{\beta}{2} \sum_i Z_i}(|+++...+\ra +|---...-\ra)
 \end{equation}
up to a normalization factor. In particular,  note that the above
state is the ground state of our model even if we remove the
non-local term, $B$, from the Hamiltonian. To see this better,
consider a local quantum Hamiltonian without the non-local term.
We will then have another (degenerate) ground state given by:
 \begin{equation}\label{2}
 |G'(\beta)\ra = e^{\frac{\beta}{2} \sum_i Z_i}(|+++...+\ra -|---...-\ra)
 \end{equation}
up to a the normalization factor. On the other hand, using the
local quantum Hamiltonian, we see that the ground state in the
limit of $\beta \rightarrow \infty$ must be the state
$|000...0\ra$. However, one can check that the state (\ref{1})
goes to $|00...0\ra$ while the state  (\ref{2}) does not satisfy
this condition. Accordingly, a sector of the ground subspace
($|G(\beta)\ra$) is important for determining the phase transition
point in a sense that it is the ground state for \emph{all} values
of $\beta$.

Now, our result shows that a symmetry breaking phase transition
occurs at infinite value of $\beta$ in a sense that the ground
state is a non-degenerate magnetized state $|000...0\ra$ at
infinite value of $\beta$ which has a $Z_2$ symmetry and it breaks
to a degenerate space for any other finite value of $\beta$. In
other words, both states ($|G\ra$ and $|G' \ra$) are ground states
of the local quantum Hamiltonian at any finite value of $\beta$
and also any arbitrary superposition of them. Specifically, one
can check that the following two quantum states are also ground
states of the local quantum Hamiltonian, up to a normalization
factor, for finite values of $\beta$:
$$|G_1 (\beta)\ra =e^{\frac{\beta}{2} \sum_i Z_i}|+++...+\ra $$
 \begin{equation}\label{}
 |G_ 2 (\beta)\ra = e^{\frac{\beta}{2} \sum_i Z_i}|---...-\ra
 \end{equation}
It is clear that the above quantum states do not have a $Z_2$
symmetry, which means that in the thermodynamic limit the $Z_2$
symmetry of the system breaks as the system selects one of the
above quantum states. Such a phase transition occurs as the system
starts from the symmetric state of $|000...0\ra$ and then the
$\beta \rightarrow \infty$ is taken. We have therefore shown that
in such quantum model there is no topological phase transition but
a symmetry-breaking phase transition which occurs at infinite
$\beta$. This means that the GHZ state does not have a topological
order but it is a symmetry-breaking quantum phase.

\begin{figure}[t]
\centering
\includegraphics[width=8cm,height=4cm,angle=0]{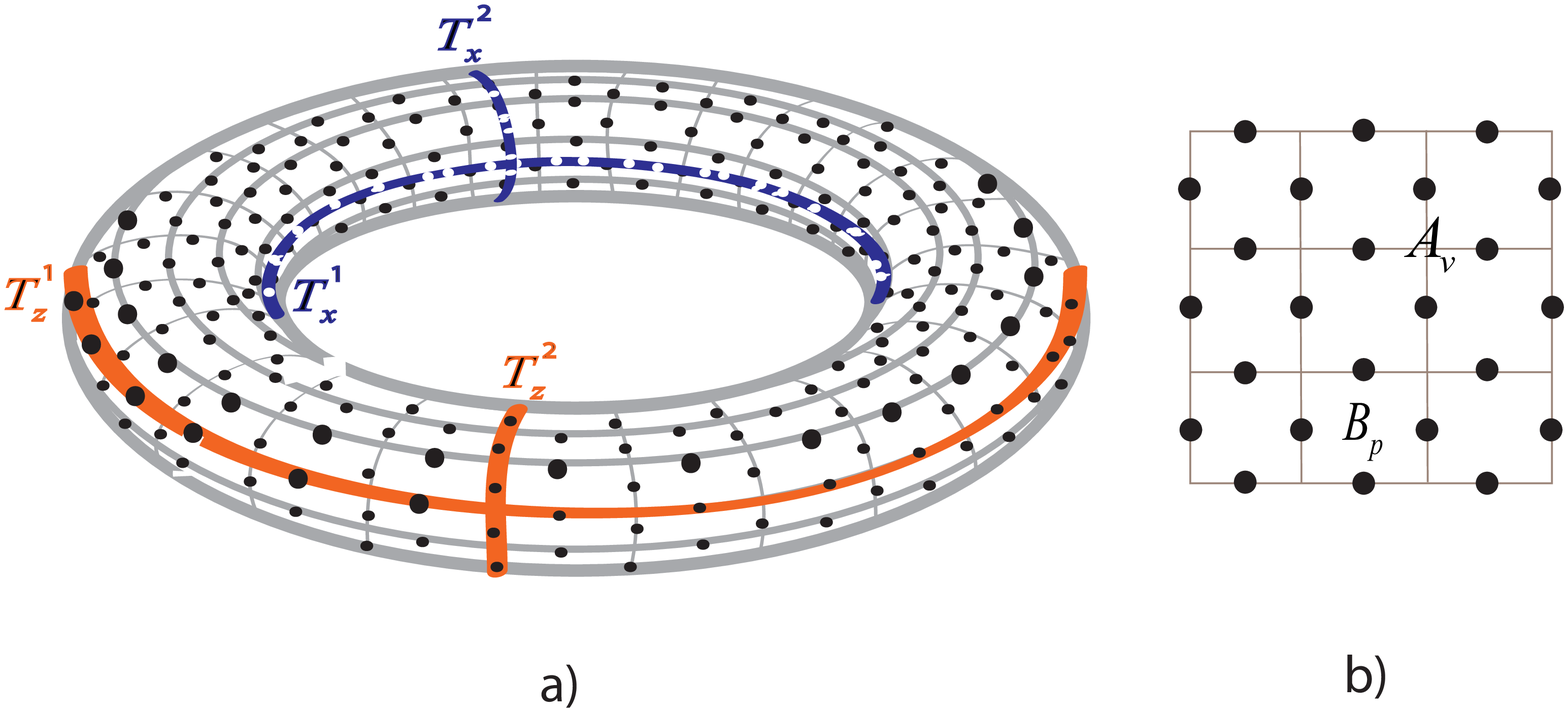}
\caption{a) Toric code is defined on a torus where qubits live on
edges of a square lattice. Corresponding to non-contractible loops
around the torus, non-local $X$ type and $Z$ type operators are
defined.  b) $X$-type and $Z$-type stabilizers are defined
corresponding to vertices and plaquettes of the lattice,
respectively. } \label{aks}
\end{figure}

As the second example, we consider two-dimensional Ising model on
a square lattice. As it has been shown in Ref. \cite{zarei18},
such model should be mapped to a Toric code state on the same
square lattice. Toric code model is defined on a torus, see
Fig.(\ref{aks}), where $X$-type stabilizers of $A_v =\prod_{i\in
v} X_i$ are defined corresponding to each vertex of the lattice
and $Z$-type stabilizers of $B_p =\prod_{i \in \partial p} Z_i$
are defined corresponding to each plaquette of the lattice. The
Hamiltonian describing toric code model is in the form of $H_1
=-\sum_{p}B_p -\sum_{v} A_v$ which has a four-fold degenerate
ground state. One of the ground states is in the following form up
to a normalization factor:
\begin{equation}
|\psi \ra =\prod_{v}(1+A_v )|00...0\ra
\end{equation}
where $|0\ra$ in the positive eigenstate of the Pauli operator
$Z$. Other degenerate ground states can also be constructed by two
non-local $X$-type operators corresponding to two non-trivial
loops around the torus, which we denote by $T_x ^1$ and $T_x ^2$
as shown in Fig.(\ref{aks}). Then, the four degenerate ground
states will be in the following form:
\begin{equation}
|\psi _{ij} \ra =(T_x ^1 )^i (T_x ^2)^j |\psi \ra
\end{equation}
where $i,j =\{0,1\}$. On the other hand, the above states can be
distinguished from each other by two non-local $Z$-type operators
$T_z ^1$ and $T_z ^2$ corresponding to non-trivial loops around
the torus, see Fig.(\ref{aks}). Because of anti-commutation
relations between non-local $X$-type and $Z$-type operators, it is
shown that effect of $T_z ^1$ and $T_z ^2$ in the ground states of
the toric code model will be in the following form:
\begin{equation}
T_z ^1 |\psi _{ij} \ra =(-1)^j |\psi _{ij} \ra ~~,~~T_z ^2 |\psi _{ij} \ra =(-1)^i |\psi _{ij} \ra
\end{equation}

Therefore, it means that $|\psi _{00} \ra =|\psi \ra$ is also
stabilized by $T_z ^1$ and $T_z ^2$. Now, turning to the
Hamiltonian of the Toric code model $H_1$, we add operators $T_z
^1$ and $T_z ^2$ to the $H_1$ to have a new Hamiltonian in the
form of $H_2 =-\sum_{p}B_p -\sum_{v} A_v -T_z ^1 -T_z ^2$. It is
clear that such a Hamiltonian has a unique ground state of
$|\psi_{00}\ra$ and there is no degeneracy for the above
Hamiltonian.

Next, the quantum CSSM model, according to definition in
Eq.{\ref{qm}, will be constructed by adding a term in the form of
$\prod_{i \in \partial p} e^{-\beta Z_i}$ to the $H_2$. On the
other hand, according to our procedure, the ground state fidelity
of such model will be mapped to heat capacity of a two dimensional
Ising model. Consequently, since the classical two-dimensional
Ising model has a phase transition in a finite temperature, the
above corresponding CSS model will show a quantum phase transition
at a finite $\beta$. This is in agreement with the fact that Toric
code state has topological order. The quantum phase transition is
not a symmetry breaking one, but a topological phase transition.

Furthermore, this also shows why we use non-local $Z$-type stabilizers
in $H_2$. To this end, note that although the Hamiltonian $H_2$ is
non-local, it has a common ground state with the initial
Hamiltonian $H_1$. On the other hand, since topological order is a
property of the wave function, both models microscopically
describe a topological order in their ground state wave function.
Next, when we add the perturbation in the exponential form of
$\sum_{v}e^{\beta \sum_{i\in v}Z_i}$ to $H_1$ and $H_2$, the state
$|G(\beta)\ra$ is still the ground state of both perturbed
Hamiltonians in a sense that for $\beta =0$, $|G(\beta)\ra$ is
equal to $|\psi_{00}\ra$ and for $\beta \rightarrow \infty$, it is
equal to magnetized state $|00...0\ra$. In other words, even for
degenerate model of $H_1$, the quantum phase transition occurs in
the sector of the Hilbert space which includes $|\psi _{00} \ra$.
It means that the above quantum phase transition is not related to
the degeneracy of the ground state. It is for this reason that we
consider non-local terms in the Hamiltonian in a sense that we are
sure that the degeneracy does not play any role and there is no
symmetry breaking phase transition in our CSSM models at a finite
$\beta$.

\end{document}